\documentclass[iop]{emulateapj}
\usepackage{apjfonts}
\usepackage{graphicx}
\usepackage[breaklinks,colorlinks,urlcolor=blue,citecolor=blue,linkcolor=blue]{hyperref}
\journalinfo{The Astrophysical Journal, 2017, in press}

\shorttitle{Mass Loss Rates from Coronal Mass Ejections}
\shortauthors{S.\  R.\  Cranmer}

\begin{document}

\title{Mass Loss Rates from Coronal Mass Ejections:
A Predictive Theoretical Model for Solar-Type Stars}

\author{Steven R. Cranmer}
\affil{Department of Astrophysical and Planetary Sciences,
Laboratory for Atmospheric and Space Physics,
University of Colorado, Boulder, CO 80309, USA}

\begin{abstract}
Coronal mass ejections (CMEs) are eruptive events that cause a
solar-type star to shed mass and magnetic flux.
CMEs tend to occur together with flares, radio storms, and bursts of
energetic particles.
On the Sun, CME-related mass loss is roughly an order of magnitude
less intense than that of the background solar wind.
However, on other types of stars, CMEs have been proposed to
carry away much more mass and energy than the time-steady wind.
Earlier papers have used observed correlations between solar CMEs
and flare energies, in combination with stellar flare observations,
to estimate stellar CME rates.
This paper sidesteps flares and attempts to calibrate a more fundamental
correlation between surface-averaged magnetic fluxes and CME properties.
For the Sun, there exists a power-law relationship between the magnetic
filling factor and the CME kinetic energy flux,
and it is generalized for use on other stars.
An example prediction of the time evolution of wind/CME mass-loss
rates for a solar-mass star is given.
A key result is that for ages younger than about 1 Gyr (i.e., activity
levels only slightly higher than the present-day Sun), the CME mass
loss exceeds that of the time-steady wind.
At younger ages, CMEs carry 10--100 times more mass than the wind,
and such high rates may be powerful enough to dispel circumstellar
disks and affect the habitability of nearby planets.
The cumulative CME mass lost by the young Sun may have been as much
as 1\% of a solar mass.
\end{abstract}

\keywords{%
stars: activity ---
stars: mass-loss ---
stars: winds, outflows ---
Sun: corona ---
Sun: coronal mass ejections (CMEs) ---
Sun: evolution}

\section{Introduction}
\label{sec:intro}

Stars never seem to be in a purely static state of mass conservation.
If they are not accreting gas from the surrounding interstellar medium,
they tend to be losing mass in the form of either a quasi-steady
stellar wind \citep{LC99,RO16} or in episodic bursts driven by a range
of possible instabilities.
Stellar mass loss has a significant impact on stellar and planetary
evolution, and also on the larger-scale evolution of gas and dust in
galaxies \citep[e.g.,][]{Wi00,OC09,Lm12,Se14}.
The Sun, in addition to having a ubiquitous steady wind, undergoes
frequent coronal mass ejections (CMEs).
These are magnetically driven eruptions of plasma and electromagnetic
energy that remain coherent even far into the outer heliosphere
\citep{Lo01,vD05,Fb06,Vr08,Sj09,Ch11,WH12,Sm15}.
CME mass loss from other stars may be able to explain observed
variability in debris disks \citep{Os13} and could be powerful enough
to strip away the atmospheres of of otherwise habitable exoplanets
\citep{Lm07,Ky16}.

Signatures of time-steady stellar winds have been detected from
stars with a wide range of properties
\citep[e.g.,][]{dJ88,Wo06,Pu08}.
However, observational evidence for intermittent mass loss is still
elusive.  There are multiple measurements of time-variable outflows
from cool stars \citep{Ho90,Cu94,FS04,Lz11,Du14,Ve16,Ko17},
but it is not yet clear whether these should be interpreted as
magnetically driven events analogous to solar CMEs.
It is possible that observing the extrasolar equivalents of
Type~II radio bursts \citep{Cy16} or the polarization signatures
of off-limb prominences \citep{Fe17} could be promising avenues toward
the goal of definitive exo-CME detection.

Even for the well-observed case of solar CMEs, many questions remain
about how they originate and evolve.
Most CMEs appear to involve the unstable expansion of twisted
``flux ropes'' (or some kind of highly non-potential, current-carrying
magnetic field) into the heliosphere.
Past attempts to winnow down the list of proposed eruption processes
have depended crucially on knowledge about how the magnetic energy is
stored in the non-potential regions in and around the flux rope---see,
e.g., evidence presented by \citet{An99} and \citet{St01} against the
\citet{Hi74} tether-cutting scenario, or the evidence presented by
\citet{Sc10} against the flux injection hypothesis of \citet{CK03}.
The true connections between temporal {\em events} such as CMEs and
flares, {\em structures} such as prominences, flux ropes, and shocks,
and {\em physical processes} such as reconnection, turbulence, and
magnetohydrodynamic (MHD) instabilities are not yet understood.

Recent progress has been made in estimating stellar CME properties by
using the Sun to normalize a correlation between the ultraviolet and X-ray
energy released in flares with CME masses and kinetic energies
\citep{Aa11,Aa12,Dr13,OW15,Tk16}.
However, this may not be the most natural way to follow the dominant
energy budget in these systems.
On the Sun, eruptive CMEs seem to be associated only with a subset
of all X-ray flares \citep[e.g.,][]{NA04,ZL05}.
Also, despite some strong correlations between stored magnetic energy
and coronal X-ray emission \citep{Pv03,Hz15}, the magnetic energy that
is released in the form of flare radiation tends to be a very small
fraction of the total.
Thus, it may make more sense to use magnetic energy fluxes than
it would to use flare emission as the primary scaling variable for
stellar CME properties.

The objective of this paper is to extend an existing semi-analytic model
of stellar wind mass-loss rates \citep{CS11} to also predict CME
mass-loss rates.
Both models assume the overall level of magnetic activity is
related to the dimensionless {\em filling factor} of strong-field
regions on the stellar surface, and that the filling factor is
correlated with the star's rotation rate (i.e., the Rossby number).
Section \ref{sec:empcorr} of this paper shows how the Sun can be used
to calibrate the correlations between magnetic filling factors and
kinetic energy losses in wind/CME flows.
Section \ref{sec:model} makes use of these correlations in building a
CME mass loss model, and illustrates it with a prediction for the time 
evolution of mass-loss rates for a one solar mass ($M_{\odot}$) star.
Lastly, Section \ref{sec:conc} summarizes the results, discusses
some wider implications of this work, and suggests future improvements.

\section{Empirical Correlations with Magnetic Flux}
\label{sec:empcorr}

The main conjecture of this paper is that a cool star's magnetic
field strength is a primary factor in determining its time-averaged
mass loss in the form of steady wind and bursty CMEs.
The Sun's observed 11-year solar-cycle variability will be used to
help determine the relationship between the kinetic energy content
in the two types of outflow (Section \ref{sec:empcorr:mdot}) and
the available magnetic energy (Section \ref{sec:empcorr:fstar}).
The correlations themselves are discussed in
Section \ref{sec:empcorr:corr}.

\subsection{Mass Loss Rates and Kinetic Energy Fluxes}
\label{sec:empcorr:mdot}

For the time-steady wind, the most appropriate solar quantity to
compare with stellar observations would be the
total sphere-averaged mass-loss rate
(i.e., accounting for all $4\pi$ steradians around the Sun).
However, such complete sampling is usually not available for solar data.
There is a large historical database of in~situ measurements in the
ecliptic plane, but {\em Ulysses} showed these are not necessarily
representative of the higher latitudes \citep{Go96}.
\citet{Wa98} produced a model-based proxy for $\dot{M}_{\rm wind}$,
the sphere-averaged solar wind mass-loss rate,
for solar cycles 21 and 22.
This model was based on known correlations between the wind speed and
density at 1~AU with the superradial expansion rate of open magnetic
flux tubes.
The latter can be reconstructed globally from rotation-averaged
magnetograms \citep[e.g.,][]{WS90}.

\begin{figure}
\epsscale{1.193}
\plotone{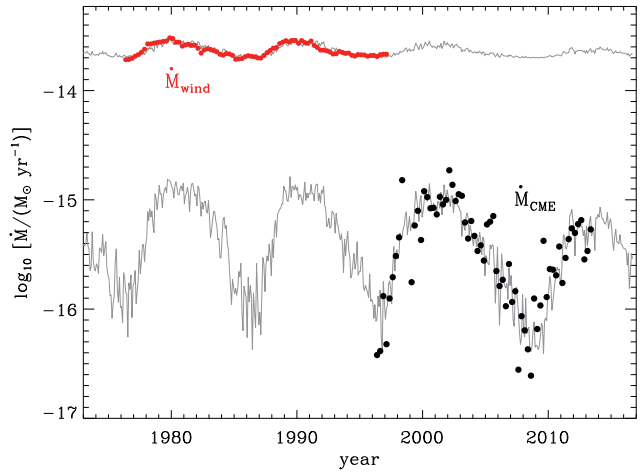}
\caption{Estimated time dependence of total rates of mass loss for the
time-steady solar wind \citep{Wa98}
(red points) and for solar CMEs
from the CDAW database (black points).
Also shown are approximate correlations between the mass-loss rates and
specified functions of sunspot number (gray curves),
given in Equations (\ref{eq:ssnwind})--(\ref{eq:ssncme}).
\label{fig01}}
\end{figure}

Figure \ref{fig01} shows the time dependence of the \citet{Wa98}
derivation of $\dot{M}_{\rm wind}$ in units of $M_{\odot}$~yr$^{-1}$.
At solar maximum there is a tendency for the Sun to produce slow and
dense wind streams at all latitudes, as opposed to the tenuous
high-speed streams that fill most of the heliospheric volume at solar
minimum.
Because the relative density increases in slow streams are slightly
larger than the relative decreases in wind speed, the sphere-averaged
mass-loss rate appears to be roughly 50\% larger at solar maximum.
For comparision,
Figure \ref{fig01} also shows an approximate correlation with
sunspot number,
\begin{equation}
  \dot{M}_{\rm wind} \, \approx \, 3.5 \times 10^{-17}
  \left( S + 570 \right) \,\,\, M_{\odot} \, \mbox{yr}^{-1}
  \label{eq:ssnwind}
\end{equation}
where $S$ is the recently revised month-averaged sunspot number from the
World Data Center (WDC) Sunspot Index and Long-term Solar Observations
(SILSO) program \citep[see, e.g.,][]{CL16}.
The correlation coefficient between the data and the above
fitting function is high (0.884), but it is likely that better fits
can be found with additional variation of the functional form.

It is possible to estimate the sphere-averaged CME mass-loss rate
$\dot{M}_{\rm cme}$ from visible-light coronagraph measurements.
Dense CMEs show up as bright features due to the fact that optically thin
Thomson scattering is linearly proportional to the line-of-sight
integrated electron density.
Thus, coronagraphs can be used to measure CME masses in a large fraction
(but not 100\%) of the volume of the extended corona and inner heliosphere.

Figure \ref{fig01} shows a reconstruction of the time-variable CME
mass-loss rate, assembled from the CME catalog maintained by NASA's
Coordinated Data Analysis Workshop (CDAW) data center
\citep[see, e.g.,][]{Ya04,Go09}.
The most recent version of the catalog contains CME records from
1996 January to 2015 February.
Out of the initial list of 25161 events, we eliminated all CMEs with
marginal detections (i.e., labeled ``poor'' and ``very poor'') and
neglected all events without tabulated masses.\footnote{%
There were 7199 events (29\% of the total) labeled ``poor''
and 6481 events (26\%) labeled ``very poor.''
There were 13706 events (54\%) without tabulated masses, and many
of these were also labeled either poor or very poor.
Most of these marginal detections seem to represent the low end
of the CME mass distribution.}
This left a subset of 6379 CMEs between 1996 March and 2013 June.

An initial summation of these masses yielded values of $\dot{M}_{\rm cme}$
that were decidedly low in comparison to the existing CME literature.
At solar maximum, the CDAW data seemed to indicate that CMEs contribute
only about 3\% of the background solar wind mass flux.
However, earlier studies of both coronagraphic and in~situ data
\citep[e.g.,][]{Hi77,Ho85,WH94} found this number to be more like
10\% to 15\%.  There are several possible reasons for such a discrepancy:
\begin{enumerate}
\item
If even a fraction of the marginal CDAW events---which we removed
from consideration---represented true CMEs, the initial summation would
have yielded a value too low in comparison with the actual CME mass flux.
The tabulated masses for these events, when they are given in
the database, are likely to be highly uncertain.
Nevertheless, including those masses for poor and very poor events
would have increased the total CME mass in the entire database by only
19\%, from $1.051 \times 10^{19}$~g (without the marginal events) to
$1.248 \times 10^{19}$~g (with those events).
\item
Standard methods of computing CME masses from coronagraph data have been
found to underestimate the masses of Earth-directed ``halo'' events,
and in some cases those CMEs could be hidden entirely behind the
instrument's occulter and thus missed completely.
\citet{Ho85} estimated that coronagraph-derived mass-loss rates may be
too low by factors of 2 to 3, mainly due to these undetected events.
\citet{Bu04} found that ``non-limb'' CMEs away from the plane of the
sky may have significantly underestimated masses, too.
\item
CME mass flux estimates from the 1970s and 1980s may be overestimates.
\citet{Vo10,Vo11} noted that older visible-light instruments were
generally less sensitive than modern CCD-based instruments.
Thus, they were more apt to detect only the most massive CMEs.
If those events were counted as ``typical,'' they may have contributed
to larger estimates of the mean mass flux than would have been obtained
from a more accurate distribution of strong and weak events.
\item
Despite the above issues, \citet{Vo10,Vo11} concluded that there may
have been a true decline in CME mass flux from the 1970s--1980s to the
1990s--2000s.
\end{enumerate}

Thus, to account for some combination of the possible underestimation
effects listed above, the mass of each CME in the reduced database
of 6379 events was multiplied by a constant factor of 1.5.
This value may still be too low---especially if the time-averaged mass
fluxes are too low by factors of 2 to 3 as suggested by
\cite{Ho85} and \citet{Vo10,Vo11}---but we did not want to stray too
far from the straightforward numbers given in the CDAW database.
The data points in Figure \ref{fig01} show the result of this
adjustment.
Each point is the sum of the CME masses in successive three-month
intervals, starting with 1996.25--1996.50 and extending to
2013.25--2013.50.
Mean rates of mass loss were computed by multiplying each summed mass
by a factor of $1.5/\Delta t$, where $\Delta t = 0.25$~yr.
As is well-known, the solar cycle dependence in $\dot{M}_{\rm cme}$
is much stronger than for the background wind's mass loss.
The data points are plotted on top of an approximate
correlation with sunspot number,
\begin{equation}
  \dot{M}_{\rm cme} \, \approx \, 5.6 \times 10^{-18}
  \left( S + 7 \right) \,\,\, M_{\odot} \, \mbox{yr}^{-1}
  \label{eq:ssncme}
\end{equation}
and the correlation coefficient between the data and the
fitting function (0.812) is only slightly lower than that of the wind.
It should be noted that Equations (\ref{eq:ssnwind}) and
(\ref{eq:ssncme}) were produced for almost completely non-overlapping
time periods.
Also, if \citet{Vo10,Vo11} were correct in their conjecture about a
true secular decrease by a factor of 2 from 1970 to 2000, the above fit
would not reproduce it.
Nevertheless, Equations (\ref{eq:ssnwind}) and (\ref{eq:ssncme})
are not used in the remainder of this paper and are presented only
as hints for future study.

The mass-loss rates given above describe time-averaged bulk outflows
of plasma from the stellar surface.
A more physically meaningful way of expressing that outflow is by
defining the mean kinetic energy flux $F$, which for a given component
of the outflow can be expressed as
\begin{equation}
  F(r) \, = \, \frac{1}{2} \rho u^3 \, = \, \frac{\dot{M} u^2}{8\pi r^2}
  \label{eq:Fkin}
\end{equation}
where the mass density $\rho$ and radial flow speed $u$ are both functions
of radial distance $r$
\citep[see, e.g.,][]{Hm82,HL95,Le82,Wi88,SM03}.

The kinetic energy flux is not always the largest term in the energy
budget of individual ``parcels'' of solar wind \citep{LC12} or
CME \citep{Mu11} plasma, but it serves well to quantify the overall
amount of material ejected in this form.
The assumption of time-steady outflow is highlighted in the last
expression in Equation (\ref{eq:Fkin}), which presumes mass conservation,
\begin{equation}
  \dot{M} \, = \, 4\pi \rho u r^2 \,\,\, ,
\end{equation}
as well as spherical symmetry far from the star.
As a function of radial distance, the energy flux $F(r)$ first increases
in the corona (as the flow accelerates), then it decreases in the
heliosphere (as $u$ approaches a constant value and the $1/r^2$ term
dominates).
Because the goal of this paper is to relate these kinetic energy fluxes
with magnetic energy fluxes (related to stellar activity and coronal
heating), it simplifies matters to define a fiducial value of $F$.
Thus, for convenience, the numerator of Equation (\ref{eq:Fkin}) is
specified at a sufficient distance from the star that $u$ has reached
its typical terminal speed, but the denominator is specified
at $r = R_{\ast}$.  In other words, we define
\begin{equation}
  F_{\rm wind} \, = \, 
  \frac{\dot{M}_{\rm wind} u_{\rm wind}^2}{8\pi R_{\ast}^2} \,\, ,
  \label{eq:Fwind}
\end{equation}
\begin{equation}
  F_{\rm cme} \, = \, 
  \frac{\dot{M}_{\rm cme} u_{\rm cme}^2}{8\pi R_{\ast}^2}
  \label{eq:Fcme}
\end{equation}
where $u_{\rm wind}$ and $u_{\rm cme}$ are estimates of the asymptotic
flow speeds far from the star.
The quantities defined above do not take on the value of $F(r)$ at any
one location, but they are consistently defined as representative proxies.
Alternately, one could specify these quantities as effective
luminosities ($L_{\rm wind} = 4\pi R_{\ast}^2 F_{\rm wind}$, with a
similar definition for $L_{\rm cme}$), but either formulation captures
the major scalings with fundamental stellar properties.

\subsection{Magnetic Filling Factors}
\label{sec:empcorr:fstar}

A star's overall level of magnetic activity can be measured by the
properties of the magnetic field that intersects its photosphere.
However, a potentially more useful quantity may be the ratio of the
surface field strength to its theoretical maximum value.
This maximum value is likely to be close to the so-called
``equipartition'' field strength that represents
equal gas and magnetic pressures.
Many young and active cool stars appear to have photospheric field
strengths of the same order of magnitude as their equipartition fields
\citep{Sa96,Sa01,Rn09}.
On the Sun, the surface-averaged field tends to be several orders of
magnitude smaller than the equipartition field of $\sim$1.5~kG
(see below).
When observed at high resolution, though, much of the Sun's magnetic
field is collected into small flux tubes that themselves exhibit nearly
equipartition field strengths \citep[e.g.,][]{St73,P78}.
Thus, the Sun has a small {\em filling factor} $f_{\ast}$ of fragmented
strong-field regions.
In the remainder of this paper, we will consider $f_{\ast}$ to be
to be equivalent to the ratio of surface-averaged field strength to the
equipartition field strength.

\begin{figure}
\epsscale{1.193}
\plotone{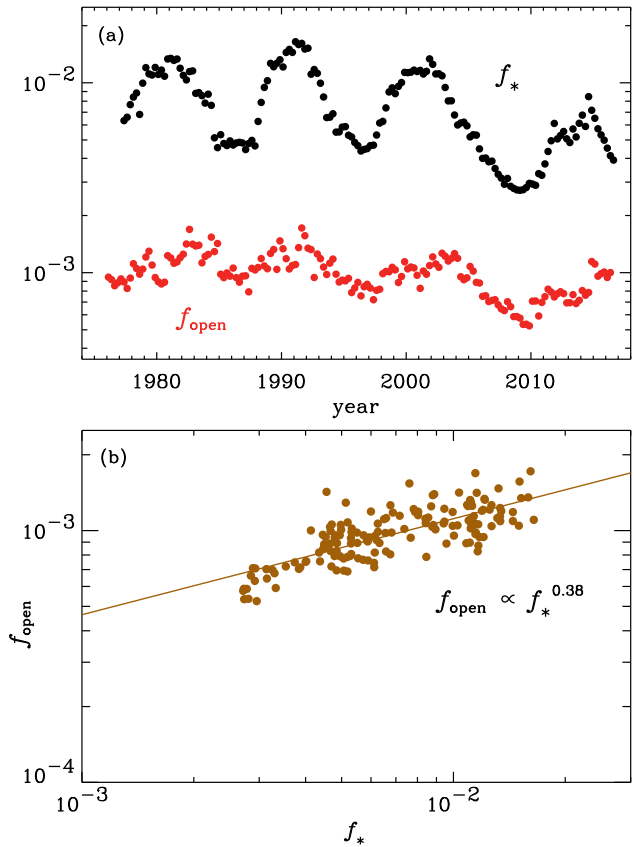}
\caption{(a) Time dependence of the Sun's total magnetic filling factor
$f_{\ast}$ (black points) and the filling factor of open magnetic
flux $f_{\rm open}$ (red points).
(b) A point-by-point comparison between $f_{\ast}$ and $f_{\rm open}$
shows that a power-law correlation reproduces much of the observed
solar-cycle variability.
\label{fig02}}
\end{figure}

\citet{CS11} followed the implications of assuming a causal relationship
between $f_{\ast}$ and the time-steady stellar wind mass-loss rate.
In order to extend that work to CME mass loss, we need to distinguish
between the total filling factor $f_{\ast}$---which accounts for all
magnetic fields, no matter their topology---and the filling
factor of open flux $f_{\rm open}$ that counts only the field lines
that connect the photosphere to the outflowing stellar wind.
The latter quantity was used exclusively by \citet{CS11}, but CME rates
appear to depend more on the non-potential magnetic activity in
closed-field active regions.
The latter should scale more like $f_{\ast}$ than $f_{\rm open}$
in environments where the total magnetic energy is dominated by these
small-scale regions.
For the present-day Sun, $f_{\ast}$ is usually about a factor of ten
higher than $f_{\rm open}$.
Figure \ref{fig02}(a) illustrates their inferred dependences on the
solar cycle \citep[see also][]{WS02}.

The solar values of $f_{\ast}$ were computed from spatially averaged
surface magnetic flux densities $|B|_{\rm av}$ from two observational
databases.
From 1977 to 2003, full-disk measurements from the National Solar
Observatory (NSO) Kitt Peak Vacuum Telescope \citep{Li76,Jn92} were used.
From 2003 to 2016, these data were supplanted by the Vector
SpectroMagnetograph instrument of the Synoptic Optical Long-term
Investigations of the Sun (SOLIS) facility \citep{Kc03,Hn09}.
In a similar manner to the $\dot{M}$ data shown in Figure \ref{fig01},
the magnetic field data were binned into 0.25~yr averages.
The filling factor $f_{\ast}$ was determined by assuming
$|B|_{\rm av} = f_{\ast} B_{\ast}$.
A fiducial value of $B_{\ast} = 1.5$~kG was used, which is slightly
higher than the standard equipartition field strength of $\sim$1.4~kG;
see Section 2.1 of \citet{CS11}.

The filling factor of open flux $f_{\rm open}$ is not as straightforward
to measure as is $f_{\ast}$.
Ideally, direct measurements of the large-scale coronal magnetic field
would be needed to distinguish between open and closed regions, but those
are not available.
In~situ magnetic field measurements at 1~AU can be used, but those
are usually limited to a single vantage point near the Earth.
However, {\em Ulysses} showed that the radial magnetic flux $r^2 B_r$
tends to be reasonably constant throughout the high- and low-latitude
heliosphere \citep{SB95}.
The open field is believed to expand laterally in the low-$\beta$ (i.e.,
magnetic pressure dominated) corona, and neighboring flux tubes eventually
reach transverse magnetic-pressure equilibrium.
Thus, the open flux wants to ``evenly'' fill the heliospheric volume.
The value of the radial flux changes with solar cycle
\citep[e.g.,][]{SC07,SB08}, but measuring it at one location in the
ecliptic seems to be an adequate proxy for its mean value taken over
all $4\pi$ of solid angle.

Thus, the time series of $f_{\rm open}$ values shown in
Figure \ref{fig02}(a) was constructed from radial magnetic field
strengths at 1~AU taken from the OMNI data set \citep{KP05}.
OMNI collects in~situ solar wind measurements from 18 different
in-ecliptic spacecraft, assembles them into a coherent and validated
database, and distributes it in a range of formats at the Space Physics
Data Facility (SPDF) of NASA's Goddard Space Flight Center.
Hour-averaged OMNI measurements were projected back to the solar surface
assuming radial flux conservation, and the values were divided by the
same equipartition field strength $B_{\ast}$ discussed above to
obtain $f_{\rm open}$.
To maintain continuity with the $f_{\ast}$ data, mean values of
$f_{\rm open}$ were constructed for each of the three-month periods
between 1976.00 and 2016.50.

As expected, $f_{\rm open}$ is smaller than $f_{\ast}$ by about an
order of magnitude, but the difference is largest at solar maximum.
Because both filling factors vary in phase with the solar cycle,
it is possible to correlate them with one another.
Figure \ref{fig02}(b) shows a least-squares power-law fit between
the two quantities, with
\begin{equation}
  f_{\rm open} \, \approx \, 0.00645 \, f_{\ast}^{0.3813}
  \label{eq:fopenfit}
\end{equation}
indicating that the relative solar-cycle dependence of $f_{\rm open}$
is significantly muted in comparison with that of $f_{\ast}$.

\subsection{Correlations with Kinetic Energy Flux Efficiencies}
\label{sec:empcorr:corr}

As stated above, the main idea of this paper is to explore the
implications of a correlation between magnetic fields and the kinetic
energy fluxes of stellar wind/CME outflows.
A major assumption is that the Sun's limited range of variation
in quantities such as $f_{\ast}$ and $F_{\rm cme}$ over the activity
cycle can be extrapolated to other stars.
Specifically, we look for relationships between the
filling factor $f_{\ast}$ and a similarly dimensionless
ratio describing the fraction of available coronal energy that the
star puts into wind/CME outflow.
Those relative fractions are defined as kinetic energy flux
efficiencies ${\cal E}_{\rm wind}$ and ${\cal E}_{\rm cme}$, with
\begin{equation}
  {\cal E}_{\rm wind} = \frac{F_{\rm wind}}{f_{\ast} F_{\ast}}
  \,\,\,\,\, \mbox{and} \,\,\,\,\,
  {\cal E}_{\rm cme} = \frac{F_{\rm cme}}{f_{\ast} F_{\ast}} \,\, .
  \label{eq:effboth}
\end{equation}
In both cases, the denominator is a magnetically weighted convective
energy flux, measured at the photosphere, that is injected into the
chromosphere and corona.
The quantity $F_{\ast}$ itself is taken to be the turbulent energy flux
inside a representative strong-field flux tube.
\citet{CS11} extracted a fiducial solar value of
$F_{\ast} = 1.5 \times 10^{8}$ erg cm$^{-2}$ s$^{-1}$ from
published models of transverse kink-mode oscillations driven by the
subsurface convection \citep{Mz02}.
When the magnetic flux tubes extend above the stellar surface and
expand to fill the volume, the kink-mode waves become shear
Alfv\'{e}n waves \citep[see also][]{CvB05}.

\begin{figure}
\epsscale{1.193}
\plotone{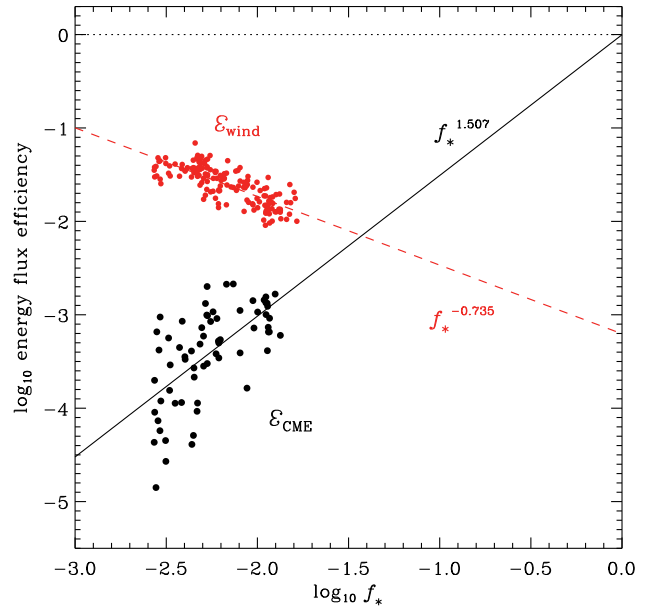}
\caption{Kinetic energy flux efficiencies for solar wind
(${\cal E}_{\rm wind}$, red points) and for CMEs
(${\cal E}_{\rm cme}$, black points) both plotted vs.\  the magnetic
filling factor $f_{\ast}$.
Power-law fits are shown for the wind (dashed red line) and for
CMEs (solid black line).
\label{fig03}}
\end{figure}

In other words, the quantity $f_{\ast} F_{\ast}$ is assumed to be the
maximum surface-averaged energy flux available for the heating and
energization of plasma in CMEs and the solar wind.
The efficiencies ${\cal E}_{\rm wind}$ and ${\cal E}_{\rm cme}$
describe what fractions of that maximum energy are tapped by the two
types of outflow.\footnote{%
Strictly speaking, a more consistent definition of
${\cal E}_{\rm wind}$ ought to involve dividing the flux by
$f_{\rm open} F_{\ast}$ instead of by $f_{\ast} F_{\ast}$,
as defined above.
However, $f_{\rm open}$ and $f_{\ast}$ are well correlated with one
another, so trends with the data should still be valid.
Also, the consistency between the two expressions in
Equation (\ref{eq:effboth}) makes for a clearer comparison between
these two quantities in Figure~\ref{fig03}.}
When examining the present-day solar cycle, we assume that $F_{\ast}$
remains fixed at the value given above, and thus we implicitly assume
that the dominant source of variability in wind/CME energy fluxes comes
from the variability in $f_{\ast}$.
Of course, when examining other stars or long-term stellar evolution,
it is clear that both $f_{\ast}$ and $F_{\ast}$ must vary.

Although the use of the energy flux $F_{\ast}$ was
motivated by \citet{CS11} on the basis of successful wave/turbulence
models of the solar wind, it is not completely clear that it should
be used to normalize the available energy for CMEs.
In many CME eruption models (see references in Section~\ref{sec:intro}),
the non-potential magnetic energy driving the eruption has been stored
up over timescales that can be quite long compared to the underlying
turbulent motions.
However, the CME mass loss rates discussed in this paper are averages
taken over even longer times.
After averaging over multiple eruptions (and stellar rotations),
the energy available to CMEs should scale with the overall strength
of the stellar dynamo.
It is that same dynamo that drives turbulent magnetic flux tubes up
through the surface, so normalizing the overall CME energy flux to
$f_{\ast} F_{\ast}$ may end up being a reasonable approximation.

Figure \ref{fig03} shows how the solar energy flux efficiencies
${\cal E}_{\rm wind}$ and ${\cal E}_{\rm cme}$ vary as a function of
the filling factor $f_{\ast}$.
The efficiencies were computed with the following details for each case:
\begin{enumerate}
\item
For the wind, the OMNI data at $r=1$~AU were used to compute $F(r)$ as
in Equation (\ref{eq:Fkin}).
The quantity $F_{\rm wind}$, as defined in Equation (\ref{eq:Fwind}),
was then determined by mapping the measured flux down to $r = R_{\ast}$
assuming an inverse-square radial dependence.
Although these data were taken from single-point measurements,
the kinetic energy flux has been found to be roughly constant as a
function of both latitude and solar cycle \citep[e.g.,][]{LC12}.
Thus, it is likely to be a good proxy for the corresponding
sphere-averaged quantity.
\item
For the CMEs, the same set of CDAW events that was used for
Figure \ref{fig01} was analyzed to compute $F_{\rm cme}$.
Here, however, the individual tabulated kinetic energies ($M u^2 / 2$)
were used instead of the masses.
As above, the three-month sums were multiplied by $1.5 / \Delta t$
to obtain the required time-averaged rates for input into
Equation (\ref{eq:Fcme}).
\end{enumerate}

Figure \ref{fig03} shows scatter plots of
${\cal E}_{\rm wind}$ and ${\cal E}_{\rm cme}$ versus $f_{\ast}$ for
all three-month bins in which both abscissa and ordinate values exist.
Although the Sun has only experienced about a factor of 6 variation in
$f_{\ast}$ over the last few solar cycles (i.e., about 0.75 dex),
the trends appear significant enough to justify fitting formulae
that can be extrapolated to wider ranges of $f_{\ast}$.
The following least-squares power-law fits are shown in Figure \ref{fig03}:
\begin{equation}
  {\cal E}_{\rm wind} \, \approx \, 6.256 \times 10^{-4} \,
  f_{\ast}^{-0.7351}
  \label{eq:effwind}
\end{equation}
\begin{equation}
  {\cal E}_{\rm cme} \, \approx \, 0.9955 \, f_{\ast}^{1.5066} \,\, .
  \label{eq:effcme}
\end{equation}
It is an interesting coincidence that the extrapolated value of
${\cal E}_{\rm cme}$ reaches a ``saturation'' value of 1 at nearly
the same time that $f_{\ast}$ becomes equal to 1.
This may be suggestive of meaningful physics behind the
suggestion that the turbulent flux $F_{\ast}$ is the relevant scaling
parameter for the CME energy budget.
However, if taken at face value, it also implies that at
$f_{\ast} \approx 1$ there may not be any remaining energy for coronal
heating, X-ray emission, or flare particle production.
The way stars partition the available energy into these different bins
must certainly change between the low activity levels seen on the Sun
and the saturated activity seen on other stars.
Better theoretical models are certainly needed
(see also Section \ref{sec:conc}).

Because the main results of this paper depend crucially on the
correlations shown in Figure \ref{fig03}, some additional statistical
investigation is warranted.
A linear regression analysis provides standard deviations for the
best-fitting values of the exponents in
Equations (\ref{eq:effwind})--(\ref{eq:effcme}).
At the $1\sigma$ level, the exponents are $-0.7351 \pm 0.04813$
(for the wind) and $1.5066 \pm 0.2292$ (for CMEs).
For the two plotted trends, the linear Pearson correlation coefficients 
between the data points and the fits are 0.775 (for the wind)
and 0.632 (for CMEs).
These values imply least-squares coefficients of determination
${\cal R}^2$ of 0.601 and 0.399 for the wind and CME data sets,
respectively.

The above ${\cal R}^2$ values imply that the power-law fits
``explain'' only about half of the variability of the data about their
respective mean values.  However, it is
possible to confidently reject the null hypothesis
of no correlation between $f_{\ast}$ and the efficiencies
${\cal E}_{\rm wind}$ and ${\cal E}_{\rm cme}$.
The classical Fisher--Snedecor $F$-test was applied to the two data
sets, and the resulting ratios of explained to unexplained variance
\citep{BR03} were $F=233.26$ (for the wind) and $F=43.219$ (for CMEs).
These correspond to probabilities $p$ for the null hypothesis of order 
$10^{-32}$ for the wind and $10^{-8}$ for CMEs.
Of course, the $F$-test does not rule out models other than the
ones given by Equations (\ref{eq:effwind})--(\ref{eq:effcme}), but
it does indicate that some meaningful correlation exists
\citep[see, however,][]{Pr02}.
Additional goodness-of-fit calculations would be possible if the
observational uncertainties of the data points were understood better,
but that is beyond the scope of this paper.

Figure \ref{fig03} indicates that ${\cal E}_{\rm wind}$ decreases
with increasing filling factor $f_{\ast}$.
In fact, this is exactly what \citet{CS11} predicted for time-steady
coronal winds.
Using the standard model parameters and the approximate scaling
given in Equation (45) of \citet{CS11}, the time-averaged mass-loss
rate is expected to scale as
$\dot{M}_{\rm wind} \propto f_{\rm open}^{5/7}$.
(Note that \citeauthor{CS11} used the symbol $f_{\ast}$ to refer
to the filling factor of open flux.)
Combining that with the correlation given in
Equation (\ref{eq:fopenfit}) of this paper, this is equivalent to
$\dot{M}_{\rm wind} \propto f_{\ast}^{0.2724}$.
For variations over our solar cycle, most of the fundamental
stellar parameters (e.g., $R_{\ast}$, $F_{\ast}$, and
$u_{\rm wind} \approx V_{\rm esc}$) are held fixed, so the predicted
scaling would be $F_{\rm wind} \propto f_{\ast}^{0.2724}$, and thus
${\cal E}_{\rm wind} \propto f_{\ast}^{-0.7276}$.
This exponent is extremely close to the least-squares value given
in Equation (\ref{eq:effwind}).

Lastly, it is interesting to note that the two curves in
Figure \ref{fig03} appear to cross one another when
$f_{\ast} \approx 0.037$.
This indicates that activity levels only slightly higher than the
present-day Sun's may start to show CMEs with comparable mass loss
as their time-steady winds.

\section{Evolution of a Solar-Type Star}
\label{sec:model}

The correlations noted above can be used to construct semi-empirical
predictions for both $\dot{M}_{\rm wind}$ and $\dot{M}_{\rm cme}$ as
functions of fundamental stellar parameters.
A particularly illustrative case is the evolutionary track of a star
having $M_{\ast} = 1 \, M_{\odot}$.
There is evidence that the ``young Sun'' produced a much denser and
more energetic gas outflow than it does today, which was likely to
have been important to early planetary evolution
\citep[e.g.,][]{Wo06,Gu07,Lm12,Jo15}.
Was this outflow dominated by CMEs?

\subsection{Activity-Rotation Relations}
\label{sec:model:rossby}

For many cool stars, there is a significant correlation between the
overall activity level---measured via a range of chromospheric and
coronal emission diagnostics---and the rotation rate
\citep{Ny84,Pz03,MH08,Wr11}.
As direct measurements of stellar magnetic fields have become available,
similar correlations for the surface-averaged field strength
(essentially $f_{\ast} B_{\ast}$) and $f_{\ast}$ itself have also
emerged \citep{Sa91,Sa01,MJ93,Cu98,Ma14,Fo16}.
Although there has been a great deal of work done to understand these
correlations as the manifestation of a stellar MHD dynamo
\citep[e.g.,][]{Db05,Ch10,Br15}, in this paper they are treated
as purely empirical scaling relations.
In other words, we assume $f_{\ast}$ can be specified as some function
of the stellar rotation rate.

\begin{figure}
\epsscale{1.193}
\plotone{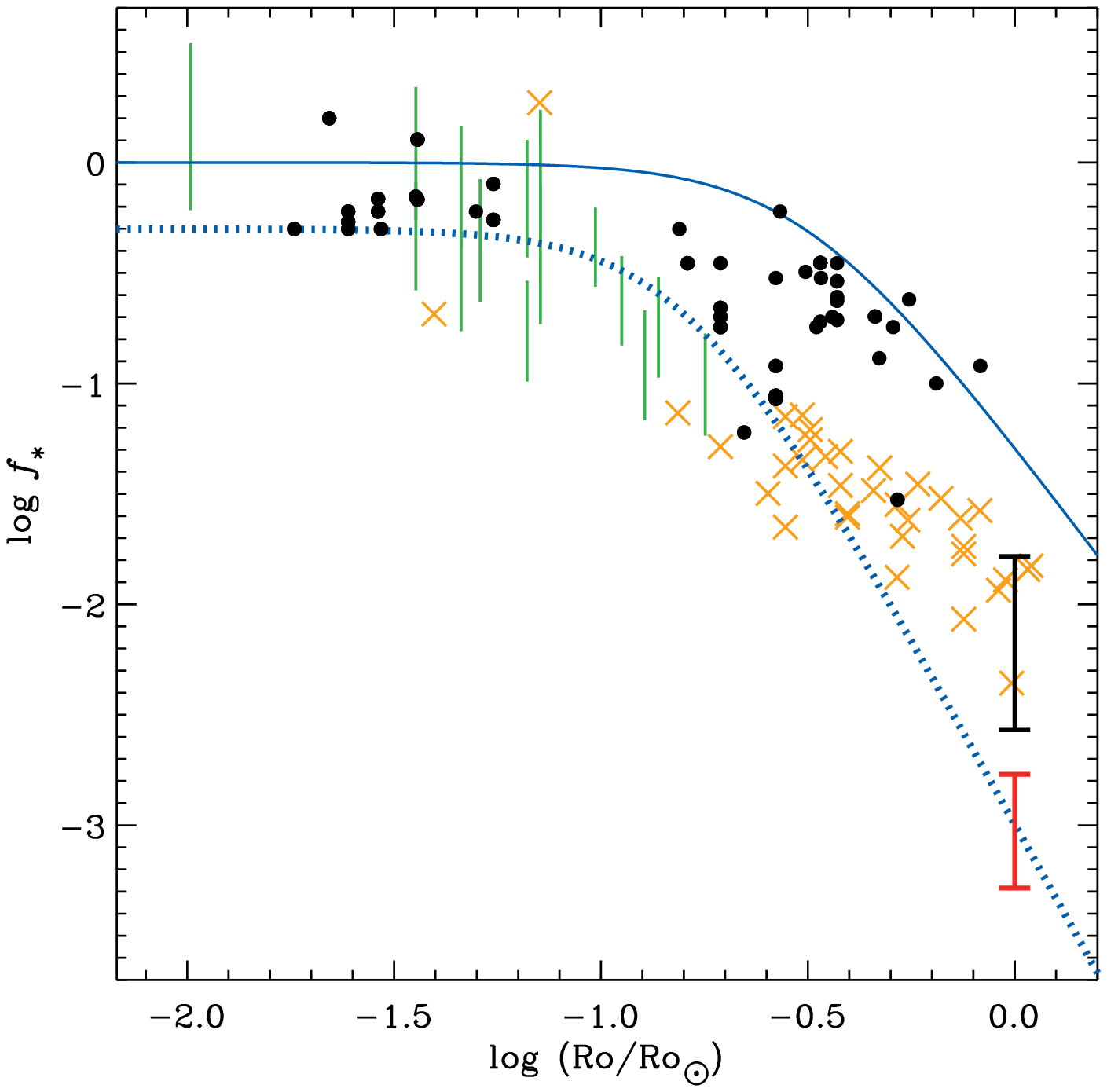}
\caption{Comparison of observationally inferred magnetic filling
factors with Rossby number, normalized by the Sun's present-day
Rossby number.  Data points are from
\citet{CS11} (black solid points),
\citet{Ma14} (orange crosses), and \citet{Fo16} (green lines).
Fitting functions for $f_{\rm min}$ (blue dotted curve) and
$f_{\rm max}$ (blue solid curve), as well as the present-day
solar-cycle variation of the Sun's $f_{\ast}$ (black strut)
and $f_{\rm open}$ (red strut) are also shown.
\label{fig04}}
\end{figure}

\citet{CS11} collected a number of $f_{\ast}$ measurements for cool stars
and parameterized their dependence on the so-called Rossby number Ro, the
ratio of the rotation period to a convective overturning timescale.
Figure \ref{fig04} shows two approximate ``envelope'' curves that
were found to encompass most of the data points.
These curves were parameterized as follows,
\begin{equation}
  f_{\rm min} \, = \, \frac{0.5}
  {[1 + (x / 0.16)^{2.6}]^{1.3}}  \,\, ,
  \label{eq:fmin}
\end{equation}
\begin{equation}
  f_{\rm max} \, = \, \frac{1}{1 + (x / 0.31)^{2.5}}
  \label{eq:fmax}
\end{equation}
where $x = \mbox{Ro} / \mbox{Ro}_{\odot}$ and the Sun's present-day
Rossby number was calibrated to be $\mbox{Ro}_{\odot}=1.96$.
See Equation (36) of
\citet{CS11} for additional details about calculating Ro and
interpreting the data.

The number of stars used by \citet{CS11} was quite limited in
comparison with more recent observational work to measure cool-star
magnetic fields.
For example, the Zeeman Doppler Imaging (ZDI) technique has provided
spatially resolved maps of vector surface fields on dozens of stars
\citep[e.g.,][]{Do12,Vi14,Se15,Fo16}.
However, one must be careful in interpreting ZDI field strengths,
because the technique is not sensitive to small-scale regions with
balanced positive and negative polarities.
Thus, ZDI field strengths are likely to be underestimates of the true
surface fluxes and filling factors.

To help quantify how flux much is missed by cool-star ZDI measurements,
\citet{Vi16} and \citet{Ja17} processed high-resolution solar data
in a similar manner as in standard ZDI analyses (i.e., they kept
the power only from low-order spherical-harmonic $\ell$ indices).
The low-order fields do a reasonable job of reconstructing the open
flux, but not the total closed flux that contributes to $f_{\ast}$.
The example computed by \citet{Vi16}, from the rising phase of solar
cycle 24, shows that limiting $\ell$ to the ZDI-sensitive values below
$\sim$10 captures only a few percent of the Sun's magnetic energy.
In this case, multiplying the ZDI surface-averaged field strength
$\langle B \rangle$ by factors of at least 5--10 would
reproduce the actual mean field strength.
However, \citet{Ja17} found that during spot-free conditions more
appropriate for solar minimum, ZDI may capture more than half of the
total magnetic flux.
This indicates a correction factor $\lesssim 2$ for low-activity times.
Thus, it appears that multiplying the ZDI-derived value of
$\langle B \rangle$ by a correction factor $>1$ goes in the right
direction, but the uncertainty on this correction factor is large.

Observational estimates of mean field strengths
$\langle B \rangle$ have been extracted from two independent surveys:
(1) \citet{Ma14} reported mean longitudinal field strengths for 67
stars; i.e., line-of-sight components of the magnetic field, averaged
over the stellar disk.
(2) \citet{Fo16} reported the strengths of individual ZDI multipole
components for 15 stars, along with the surface-averaged field
$\langle B \rangle$.
The inferred values of $\langle B \rangle$
were converted into $f_{\ast}$ by multiplying by a constant correction
factor and dividing by $B_{\ast}$ values computed from each star's
fundamental parameters, as in \citet{CS11}.
A correction factor of 7---more appropriate for solar maximum fields than
solar minimum fields---was chosen because most of these stars are more
active than the Sun, and thus are likely to be more similar to the most
active phases of the present-day solar cycle.
For the data taken from \citet{Ma14}, Figure \ref{fig04} shows only stars
that had relative uncertainties in $\langle B \rangle$ less than 10\%.
For \citet{Fo16}, we plotted each star as a vertical bar that extends
from its ZDI-derived value of $\langle B \rangle$ up to its peak value
of surface $B$ measured over rotational phase.
The Rossby numbers corresponding to these data points were taken
from their respective papers.  Recomputing them using the \citet{CS11}
method does not produce any substantial difference in the appearance
of Figure \ref{fig04}.

The corrected points from \citet{Ma14} and \citet{Fo16} fit roughly
inside the envelope defined by the $f_{\rm min}$ and $f_{\rm max}$
curves at large and small Rossby numbers, but the agreement is worse
at intermediate values.
The ZDI-derived data appear to follow a shallow power-law dependence
on Rossby number, with roughly $f_{\ast} \propto \mbox{Ro}^{-1.3}$.
This stands in contrast to the steeper limiting slopes of 
$f_{\rm max} \propto \mbox{Ro}^{-2.5}$ and
$f_{\rm min} \propto \mbox{Ro}^{-3.4}$ in the limit of slow rotation
($\mbox{Ro} \gtrsim 1$).
Of course, using a single constant correction factor of 7
for the ZDI data is not likely to be valid across a large range of
activity levels.
It is possible the different slopes could be reconciled with one another
if a more physically motivated correction procedure was used.
The implications of the different slopes on the age dependence
of CME mass-loss rate are discussed below.

\subsection{Predicted Mass Loss History}
\label{sec:model:mdot}

In order to apply the scaling relations defined above, the effective
velocities $u_{\rm wind}$ and $u_{\rm cme}$ need to be specified.
\citet{CS11} assumed that $u_{\rm wind} = V_{\rm esc}$, the surface
escape speed, which for the present-day Sun is 618 km~s$^{-1}$.
Observed CME speeds tend to be comparable to solar wind speeds,
but the solar cycle distribution is broader and more skewed to higher
values than that of $u_{\rm wind}$ \citep[see also][]{Cy00,OC04,Yu05}.
The CDAW catalog contains representative coronal values of $u_{\rm cme}$
for each event.
For the CMEs included in Figures \ref{fig01} and \ref{fig03} above,
the distribution of speeds is similarly broad and skewed as has been
reported in the literature.
The mean and median values are 449 and 402 km~s$^{-1}$, respectively,
and the standard deviation is 224 km~s$^{-1}$.
Although there is a hint of a trend with solar activity (with larger
values at solar maximum), we adopt the simple relationship
$u_{\rm cme} \approx 0.7 V_{\rm esc}$ that is consistent with the
present-day mean and median.

Assembling together Equations (\ref{eq:Fcme}), (\ref{eq:effboth}),
(\ref{eq:effcme}), and the above scaling for $u_{\rm cme}$, the
mean CME mass-loss rate is given by
\begin{equation}
  \dot{M}_{\rm cme} \, \approx \, 51.06 \left(
  \frac{F_{\ast} R_{\ast}^2}{V_{\rm esc}^2} \right) f_{\ast}^{2.5066}
  \,\,\, .
  \label{eq:mdotfull}
\end{equation}
Instead of using the age dependence implied in the limiting curves of
$f_{\rm min}$ and $f_{\rm max}$, we estimated two intermediate tracks
that agree with present-day solar minimum and maximum activity levels.
These tracks can be specified by interpolating between the two envelope
curves defined by Equations (\ref{eq:fmin}) and (\ref{eq:fmax}), with a
new effective filling factor defined as
\begin{equation}
  f_{\rm eff} \, = \, f_{\rm min}^{1-\delta} f_{\rm max}^{\delta}
  \,\, .
\end{equation}
The parameter $\delta$ indicates the fractional extent to which an
intermediate curve spans the gap between $f_{\rm min}$ (i.e., $\delta = 0$)
and $f_{\rm max}$ (i.e., $\delta = 1$).
Curves that intercept the current range of solar activity levels
correspond roughly to
$\delta = 0.30$ (present-day solar minimum $f_{\ast} \approx 0.003$) and
$\delta = 0.63$ (present-day solar maximum $f_{\ast} \approx 0.012$).

\begin{figure}
\epsscale{1.193}
\plotone{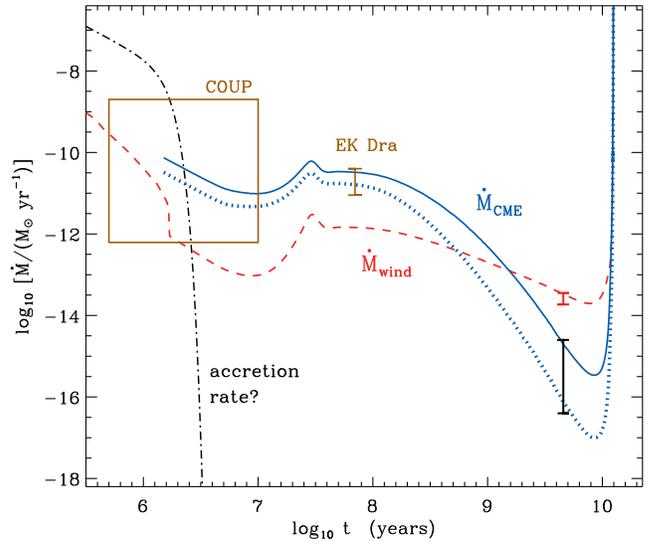}
\caption{Theoretical predictions of the mass-loss evolution of a
$1 \, M_{\odot}$ star, showing the time-steady $\dot{M}_{\rm wind}$
(red dashed curve), limiting values of $\dot{M}_{\rm cme}$ for
$\delta = 0.30$ (blue dotted curve) and $\delta = 0.63$ (blue solid
curve), and an order-of-magnitude estimate of the young-Sun accretion
rate (black dot-dashed curve).
Also shown are present-day measurements for the mass-loss rates (red
and black error bars) and flare-based estimates of $\dot{M}_{\rm cme}$
for EK Dra (brown error bar) and T~Tauri stars from the COUP database
(brown rectangle).
\label{fig05}}
\end{figure}

Figure \ref{fig05} shows the evolutionary history of steady wind and CME
mass-loss rates for a solar-mass star.
Evolutionary tracks for stellar radius, luminosity, and effective
temperature as a function of age $t$ were taken from the tabulated
$1 \, M_{\odot}$ model of \citet{Pi04}.
The rotational evolution was taken from the model of \citet{Dn10}.
These parameters were used to compute the quantities on the right-hand side
of Equation (\ref{eq:mdotfull}) using the formulae given by \citet{CS11}.
As described above, the two plotted curves for $\dot{M}_{\rm cme}$ were
computed with $\delta = 0.30$ and 0.63.

Figure \ref{fig05} also contains estimates for the time-steady solar
wind mass-loss rate $\dot{M}_{\rm wind}$ and the mean gas accretion rate
during the T~Tauri phase.
The curve for $\dot{M}_{\rm wind}$ contains two parts:
(1) for $\log t \gtrsim 6.2$, it is identical to that shown in
Figure~14 of \citet{CS11}, and
(2) for $\log t < 6.2$, the larger accretion-powered mass-loss rate
predicted by \citet{Cr08} is shown instead.\footnote{%
Because of the potential importance of accretion-driven turbulence on
the stellar surface during the classical T~Tauri phase, the predicted
CME mass loss in Figure \ref{fig05} was not extrapolated back beyond
$t \approx 1.5$ Myr.
Earlier than that, the stellar activity (and thus the CME mass-loss rate)
may be enhanced in a similar way as the time-steady wind appears to be
\citep[e.g.,][]{Cr08,Cr09}, but those effects still need to be modeled.}
The accretion rate shown in Figure \ref{fig05} is a simplistic
convolution of a power law decline of $t^{-1.5}$ at young ages
\citep[see][]{Ha98} and a rapid exponential decay that takes hold after
a few Myr (i.e., when the primordial gas disk is expected to dissipate).
The accretion rate is shown only for relative comparison with the
various predicted outflow components.

A major conclusion to be drawn from Figure \ref{fig05} is that,
at ages younger than about $\sim$1~Gyr, the CME mass loss from a
$1 \, M_{\odot}$ star exceeds that from the more steady wind that is
accelerated along large-scale open field lines.
Over the first 0.3 Gyr of a solar-mass star's lifetime (i.e.,
$\log t \lesssim 8.5$), this model predicts that $\dot{M}_{\rm cme}$
may exceed $\dot{M}_{\rm wind}$ by factors of 10 to 100.
It should also be noted that between $\sim$0.3 Gyr and the present,
$\dot{M}_{\rm cme}$ drops off roughly as $t^{-3}$ to $t^{-4}$.
This comes mainly from the $f_{\ast}^{2.5}$ dependence in
Equation (\ref{eq:mdotfull}), in combination with the rough scalings
$f_{\ast} \propto \mbox{Ro}^{-3}$ and $\mbox{Ro} \propto t^{0.5}$.
If, however, the shallower power-law $f_{\ast} \propto \mbox{Ro}^{-1.3}$
implied by the ZDI data is valid \citep{Ma14,Fo16}, this would imply
a similarly shallow age dependence of
$\dot{M}_{\rm cme} \propto t^{-1.5}$ to $t^{-2}$.
If the CME mass-loss rate is normalized at the present-day values,
this alternate scaling would mean that $\dot{M}_{\rm cme}$ would be
much lower at younger ages.

For additional observational context, Figure \ref{fig05} shows two
example flare-based inferences of $\dot{M}_{\rm cme}$ for young stars.
The large rectangle indicates ages and mass-loss uncertainty limits
for the T~Tauri stars measured by the {\em Chandra} Orion
Ultradeep Project (COUP).
The ages were estimated by \cite{PF05} and the CME mass-loss rates
were estimated by \citet{Aa12}.
Also shown is a smaller range of uncertainties for EK~Dra, a young
solar analog whose flaring has been observed extensively.
For this star, \citet{OW15} estimated the plotted range of values for
$\dot{M}_{\rm cme}$; the agreement with the model-based curves is
rather good.
This agreement may be indirect evidence in favor of the steeper
relationship between Rossby number and magnetic filling factor
($f_{\ast} \propto \mbox{Ro}^{-2.5}$ to $\mbox{Ro}^{-3.4}$) embodied
in the $f_{\rm min}$ and $f_{\rm max}$ curves in Figure \ref{fig04}.

\section{Discussion and Conclusions}
\label{sec:conc}

This paper has shown the existence of correlations between the
Sun's surface-averaged magnetic flux and the mean kinetic energy
fluxes of the solar wind and CMEs.
By framing these as correlations between dimensionless filling factors
and efficiencies, the goal was to generalize them to be able to
predict CME mass-loss rates for cool stars over a wide range
of ages and fundamental parameters.
The resulting prediction for the time evolution of a solar-mass star
with solar composition (Figure \ref{fig05}) showed good agreement with
independent estimates of $\dot{M}_{\rm cme}$ for other young solar
analogues.
During much of the first billion years of the model 1~$M_{\odot}$
star's existence, the predicted CME mass-loss rate is roughly an order
of magnitude higher than that of the time-steady wind.
The present-day reversal of that situation is facilitated by a
substantially faster time-decay for $\dot{M}_{\rm cme}$ than for
$\dot{M}_{\rm wind}$.

This work built on earlier theoretical models of time-steady solar
wind scaling relations \citep{CS11} and on empirical correlations
between CME properties and high-energy flare emissions
\citep[e.g.,][]{Aa12,Dr13,OW15}.
Although it is likely that the magnetic properties of a star are more
fundamental (i.e., they are what drive the flare and CME properties),
it is also undeniable that they are much more difficult to observe
than, say, flare light curves.
Another potentially useful observable may be total spot coverage, which
may be measurable from stellar light curves \citep{Sr09}
and has been shown---at least for individual sunspot groups---to be
correlated with flare X-ray flux \citep{Sa00}.
Thus, it would be highly beneficial to build theoretical models that
self-consistently combine all four major aspects of episodic
variability (magnetic fields, photospheric spots, flares, and CMEs).
This would allow us to validate the existing correlations and specify
how far the parameters can be extrapolated for other stars.

Predictive models of wind/CME mass loss can be used to help constrain
models of stellar rotational evolution, primordial disk depletion,
and particle ablation of planetary atmospheres.
\citet{Dr13} also suggested that CME mass loss may help explain the
so-called ``faint young Sun problem,'' in which the young Earth
appears to have had liquid water even though the Sun's luminosity
would have implied global temperatures below the freezing point
\citep{SM72,F12}.
One possible solution to the problem is that the young Sun may have
been more massive (and thus more luminous) than standard solar models
predict.
If the Sun had lost roughly 3\% to 7\% of its initial mass over its
first few Gyr \citep{SB03,MM07}, its early luminosity may have been
high enough to resolve the problem.

Although the model presented in this paper has a relatively high
cumulative mass loss due to CMEs, it does not appear to be enough to
solve the faint young Sun problem. 
The curves shown in Figure \ref{fig05} were integrated in time
between $\log t = 6.17$ and 9.66.
The mass lost by the time-steady wind is about
0.085\% of a solar mass.
The masses corresponding to the two CME curves are
0.33\% (minimum) and 0.87\% (maximum) of a solar mass.
Because these numbers are dominated by long-term behavior over timescales
of 0.1--1 Gyr, they are insensitive to the exact choice of starting age.
The active-Sun number of about 1\% agrees with the flare-based
prediction of \citet{KL14}, but it appears insufficient to raise the
young Sun's luminosity enough to solve the overall freezing problem.

Both the data and models presented in this paper can be improved
in several ways to increase the accuracy of the results.
For example, a more comprehensive accounting of coronagraph-derived
CME masses and kinetic energies should be performed in order to
reduce the existing uncertainties at the level of factors of 2--3.
Also, the derivation of magnetic filling factors $f_{\ast}$
from ZDI measurements---and reconciling these data with the
Zeeman-broadened filling factors from unpolarized stellar
spectra \citep{Sa01}---needs to be improved.
Lastly, the relatively large predicted values of $\dot{M}_{\rm cme}$
discussed above should be reconciled with some measurements that failed
to detect such high rates of mass loss \citep[e.g.,][]{LW96,Wo14}.
\citet{Dr17} suggested that on very active stars, some CMEs may be
trapped or ``stalled'' beneath large-scale magnetic loops that exert
substantial magnetic tension on underlying unstable prominences.
Their associated flares may still appear in light curves, but their
mass may be recycled back down to the coronal base.

For young T~Tauri stars still experiencing active accretion, 
\citet{Cr08,Cr09} showed that photospheric MHD turbulence---which
is likely to drive coronal activity---can be produced by two possible
mechanisms: convective motions (from below) and impacts from ``blobs''
flowing along magnetospheric accretion streams (from above).
As mentioned above, it is possible that for ages less than about
1~Myr, the second source of turbulence may cause $\dot{M}_{\rm cme}$ 
to jump up by an order of magnitude in the same way that
$\dot{M}_{\rm wind}$ does in the \citet{Cr08,Cr09} models.
This would increase $\dot{M}_{\rm cme}$ to about 10\% of the accretion
rate, which appears to be a necessary condition for enabling the
observed wind-torque spindown of solar-type stars \citep{MP05,MP08}.

A final example of the broader importance of studying CME mass loss
on active stars is their potential importance to regulating stellar
dynamos.
It has been proposed that CMEs help shed one cycle's dominant
magnetic helicity, which would otherwise build up in the convection
zone, to make room for the next cycle's opposite helicity
\citep{BF00,Lo01,Bx07}.
These insights came from studying the modern-day solar case, for
which CMEs only make up a small fraction of the mass loss.
However, in cases of CME-dominated mass loss on active stars,
the resulting dynamo may operate in qualitatively different ways
than that of the present-day Sun.

\acknowledgments

The author gratefully acknowledges
Alicia Aarnio, Avery Schiff, and Joan Burkepile
for valuable discussions.
This work was supported by the National Aeronautics and Space
Administration (NASA) under grants {NNX\-15\-AW33G} and
{NNX\-16\-AG87G}, and by the National Science Foundation (NSF)
under grants 1540094 and 1613207.
This work utilizes SOLIS data obtained by the National Solar Observatory
(NSO) Integrated Synoptic Program, managed by the NSO, which is operated
by the Association of Universities for Research in Astronomy (AURA)
under a cooperative agreement with the NSF.
Also, this paper made use of NSO's Kitt Peak magnetic data, which was
produced cooperatively by NSF/NOAO, NASA/GSFC and NOAA/SEL.
The author acknowledges use of OMNI data from NASA's Space Physics
Data Facility OMNIWeb service.
Revised sunspot number data were obtained from the WDC/SILSO program of
the Royal Observatory of Belgium, Brussels.
The LASCO CME catalog used in this paper is generated and maintained at
the CDAW Data Center by NASA and the Catholic University of America in
cooperation with the Naval Research Laboratory.  {\em SOHO} is a project
of international cooperation between ESA and NASA.

\end{document}